\documentclass[12pt]{iopart}
\usepackage{iopams}
\usepackage{amssymb}
\usepackage{xcolor}
\usepackage[titletoc]{appendix}

\usepackage{relsize}
\usepackage{cite}
\usepackage{bm}
\usepackage{changes}
\newcommand{\CQ}{{\mathcal Q}}

\newcommand{\CK}{{\mathcal K}}
\newcommand{\CM}{{\mathcal M}}

\newcommand{\CP}{{\mathcal P}}

\newcommand{\CR}{\mathcal{R}}

\newcommand{\CT}{\mathcal{T}}

\newcommand{\Ratio}{\Gamma_P}
\newcommand{\RatioU}{\Gamma}
\newcommand{\average}[1]{\left\langle #1 \right\rangle}

\newcommand{\say}[1]{`#1'}

\usepackage{hyperref}
{
 \definecolor{BLACK}{gray}{0}
 \definecolor{WHITE}{gray}{1}
 \definecolor{RED}{rgb}{1,0,0}
 \definecolor{GREEN}{rgb}{0,1,0}
\definecolor{dgreen}{rgb}{.1,.6,.1}
\definecolor{BLUE}{rgb}{0,0,1}
 \definecolor{CYAN}{cmyk}{1,0,0,0}
 \definecolor{MAGENTA}{cmyk}{0,1,0,0}
 \definecolor{YELLOW}{cmyk}{0,0,1,0}
 \definecolor{aw}{rgb}{0.2,0.5,0.75}
  }
  \hypersetup{
    colorlinks,%
    citecolor=blue,%
    filecolor=blue,%
    linkcolor=magenta,%
    urlcolor=blue
}

\definecolor{MyB}{rgb}{0.1,0.1,1.0}

\definecolor{MyGreen}{rgb}{0.0,.5,0.0}

\definecolor{MyDarkRed}{rgb}{0.7,0,0}

\begin{document}
% Local definitions
\def\beq{\begin{equation}} \def\eeq{\end{equation}}
\def\bea{\begin{eqnarray}} \def\eea{\end{eqnarray}}
\def \Scalar {S}
\def \Sc {A}
\def \VecField {\bm{V}}
\def \Vec {V}
\def \Bound {B}
\def \Vol {\mathcal{V}}
\def \norm {\mathcal{N}}
\def \heavy {\mathcal{H}}
\def \deltafun {\delta}
\newcommand\independent{\protect\mathpalette{\protect\independenT}{\perp}}
\def\independenT#1#2{\mathrel{\rlap{$#1#2$}\mkern2mu{#1#2}}}

\title[Covariance of scalar averaging and backreaction]{On the covariance of scalar averaging and backreaction in relativistic inhomogeneous cosmology}

\author{Asta Heinesen$^{1}$, Pierre Mourier$^{2}$ and Thomas Buchert$^{2}$}
\address{$^1$School of Physical \& Chemical Sciences, University of Canterbury,
Private Bag 4800, Christchurch 8140, New Zealand}
\address{$^2$Univ Lyon, Ens de Lyon, Univ Lyon1, CNRS, Centre de Recherche Astrophysique de Lyon UMR5574, F--69007, Lyon, France \\
\medskip
Emails: asta.heinesen@pg.canterbury.ac.nz and pierre.mourier@ens--lyon.fr and buchert@ens--lyon.fr}

\begin{abstract}
We introduce a generalization of the $4-$dimensional averaging window function of Gasperini, Marozzi and Veneziano
(2010) that may prove useful for a number of applications. 
The covariant nature of spatial scalar averaging schemes to address the averaging problem in relativistic cosmology is an 
important property that is implied by construction, but usually remains implicit. 
We employ here the approach of Gasperini et al. for two reasons. First,
the formalism and its generalization presented here are manifestly covariant. Second, the formalism is convenient for 
disentangling the dependencies on foliation, volume measure, and boundaries in the averaged expressions entering 
in scalar averaging schemes. These properties will prove handy for simplifying expressions, 
but also for investigating extremal foliations and for comparing averaged properties of different foliations directly.  
The proposed generalization of the window function allows for choosing the most appropriate averaging scheme for the physical problem at 
hand, and for distinguishing between the role of the foliation itself and the role of the volume measure in averaged dynamic 
equations. 
We also show that one particular window function obtained from this generalized class results in an averaging scheme corresponding 
to that of a recent investigation by Buchert, Mourier and Roy (2018) and, as a byproduct, we explicitly show that the 
general equations for backreaction derived therein are covariant. 
\end{abstract}
{\it Keywords\/}: general relativity---foliations---Lagrangian description---backreaction
%\PACS{98.80.-k, 95.36.+x, 98.80.Jk, 04.20.-q, 04.20.Cv}
%
%98.80.-k 	Cosmology (see also section 04 General relativity and gravitation; 
%				for origin and evolution of galaxies, see 98.62.Ai; 
%				for elementary particle and nuclear processes, see 95.30.Cq; 
%				for dark matter, see 95.35.+d; for dark energy, see 95.36.+x; 
%				for superclusters and large-scale structure of the Universe, see 98.65.Dx)
%95.36.+x 	Dark energy (see also 98.80.-k Cosmology)
%98.80.Jk 	Mathematical and relativistic aspects of cosmology
%04.20.-q 	Classical general relativity (see also 02.40.-k Geometry, differential geometry, and topology)
%04.20.Cv 	Fundamental problems and general formalism
%

% ==================================================

\section{Introduction} \label{sec::intro}
Cosmology is the discipline of describing overall dynamic properties of the Universe in a spatially and/or statistically averaged sense. 
For a cosmology founded on general relativistic principles, this aim is hard to obtain for at least two reasons:
\begin{itemize}
\item[(i)]
In general relativity a global and canonical notion of time is not in general expected to exist. There is no unique and general way of extending the eigentime of a world line to a global time parameter at each point in space-time. Thus, global dynamics is not easily defined since a natural \say{laboratory frame} is missing. A cosmological model would usually describe congruences of fundamental observers following source fluid flows, and would naturally attempt to build global frames based on such a family of observers. However, the identification of observer congruences in our space-time, that \say{at present day} involves a complicated hierarchy of structure, is a difficult task. Moreover, a congruence of fluid-comoving observers does not build global rest frames in presence of vorticity (expected to appear on small scales), so that alternative definitions of observers-based spatial sections may be required.
\item[(ii)]
Averages and statistical descriptions are not naturally formulated within general relativity. 
Tensor quantities are intrinsic to the tangent-space in which they live; while there are ways of mapping tensor quantities between tangent-spaces, such mappings are not unique. Furthermore, point particles as matter sources are not compatible with the formulation of general relativity. Projecting such a particle picture into a continuous space-time setting may for instance involve an extension to a curved manifold of the Newtonian procedure of coarse-graining particles in phase space by filtering the Klimontovich density and of forming appropriate moments. For these reasons statistical matter descriptions are highly involved in general relativity. 
\end{itemize}
The standard paradigm of cosmology relies on pre-assuming a statistical geometry and a corresponding matter description (disentangled from curvature degrees of freedom). 
Assuming also decoupling of scales, approximate large-scale statistical homogeneity and isotropy is used as a motivation for taking the Friedman-Lema\^itre-Robertson-Walker (FLRW) class of metrics as an idealization for the average properties of the Universe on the largest scales. However, the FLRW class of metrics assumes local isotropy which results in a homogeneous geometry on all scales, not only on the largest scales. The assumption that the FLRW geometries match the average properties does not follow from first principles.

In the field of inhomogeneous cosmology we are interested in studying the failure of the FLRW idealization as an accurate description of geometry on the largest scales, meaning the failure of it to describe the average dynamics of inhomogeneities propagating on all scales and the motions of test particles through them. 
In general relativity geometry and matter couple \textit{locally}. This core feature is missed by any large-scale description that neglects structure on small scales and only deals with coupling between an assumed large-scale geometry and averaged matter sources. 

The usual approaches to describe structure on cosmological scales involve weak field approximations around a homogeneous background. However, typical weak field argumentation in cosmology has limitations. 
It is assumed that there is a \emph{global} FLRW background metric around which the weak field is to be taken everywhere; clearly local potentials associated with most structures in our Universe are weak; the question in cosmology is what an appropriate \emph{background} is for such a weak field limit \cite{CEP}.
Moreover, even if metric perturbations are small with respect to a global FLRW metric over a spatial section of the Universe, their derivatives can be non-perturbative. This is the case for non-linear density fields (which are present at nested scales in our present-day Universe), in which case second-order derivatives of the potentials are necessarily non-linear (see, e.g. \cite{curvatureestimate}). In such cases, the expansion of the Einstein equation into a zero-order FLRW part and a first-order part breaks down, and from first principles we would not expect the FLRW field equations to be satisfied as independent equations decoupled from the dynamics of structures.

Here, we shall focus on quantifications of the non-linear backreaction of smaller scales on the large scale evolution that involves averaging of \say{local} quantities. 
We shall focus only on averaging schemes for space-time scalars as done in \cite{dust,perfectfluid},
and later generalized by many authors (see, e.g. the reviews \cite{buchertrasanen,ellisreview} and references therein).
We note that the fundamental problems in describing averaged cosmological dynamics as outlined in (i) and (ii) are not fully addressed in this form of averaging. 
In particular, the assumption of a \say{local} fluid description, where fluid elements are implicitly coarse-grained by neglecting their internal curvature degrees of freedom, is built into the Buchert equations \cite{dust,perfectfluid} (see, e.g.\cite{whatisdust}).
However, we do not assume an averaged homogeneous and isotropic fluid as a source for a large-scale statistical geometry: geometry and matter couple at the fluid resolution scale. 
The average behaviour is formulated directly from the physics at this \say{local} scale, and inhomogeneities at local scales appear explicitly in the resulting generalizations of the Friedmann equations, reflecting the non-commutativity of averaging and evolution in time. 

In this work we introduce a $4-$dimensional averaging window function that generalizes the window function presented in \cite{generalbackreac1,generalbackreac2} for integration over hypersurfaces. 
There are multiple purposes in doing so. 
First, we shall often be interested in a fluid-intrinsic averaging operation (when a fundamental fluid exists in our space-time); such intrinsic formulation will in general not be compatible with the class of window functions considered in \cite{generalbackreac1,generalbackreac2}. 
Second, the generalized scheme allows for maximal freedom in the choices of averaging domain and volume measure, while still being compact and easy to interpret. 
Covariance is built explicitly into the averaging scheme, guaranteeing that any generalization of the Buchert scheme formulated from this will be coordinate-independent by construction. 
Third, the introduction of the new window function has applications for further investigations on extremal foliations and on the dependence of averaged quantities on the foliation. Such studies are beyond the scope of this paper, but will be considered in a future paper \cite{foliationdependence}.

We are solely concerned with \emph{covariance} here; we do not consider gauge-invariance as defined in standard model perturbation theory.\footnotemark
\footnotetext{We emphasize the focus of this paper on covariant variables only, in distinction to \cite{generalbackreac1} where both covariance and standard model perturbation theory gauge invariance are discussed.}
In standard model perturbation theory the fields of interest are perturbation degrees of freedom of the space-time metric defined relative to a background metric. These fields are defined in terms of components of the metric and the background metric and \emph{do not} transform as tensors in the differential geometry definition of a tensor, i.e. they are not covariant.
This includes the Bardeen variables, which are \say{gauge-invariant} in this context, i.e. they are invariant under first-order changes of the diffeomorphism between the background manifold and the physical space-time manifold, but they are not 
$4-$scalars.

We emphasize that there is no reference to a background space-time in the context of this paper, and that we use the conventional general relativistic wording throughout. 
When referring to scalar degrees of freedom we mean quantities that do not transform under arbitrary coordinate transformations. 
When we refer to \say{gauge} degrees of freedom in this paper, this will be in the broad sense of the word, i.e. as redundant degrees of freedom in the parameterization of a physical system.

This paper is organized as follows. In section~\ref{sec:averagingscheme} we introduce the averaging scheme as formulated in terms of a covariant window function. We discuss the interpretation of the generalized adapted volume measure entering this scheme and we give examples of relevant subcases. 
In section~\ref{Commutation} we discuss the commutation rule for such an averaging operation and apply it to the conservation of regional rest mass. 
The averaged Einstein equations for a general fundamental fluid source are derived in section~\ref{sec:AEE} for a general window function, expressed in such a way that boundary terms vanish by construction, except for the average energy conservation law. 
We consider domains propagated along the fluid world lines as a special case that allow for a more transparent interpretation of the averaged equations. We conclude in section~\ref{Conclusion}.

\section{The averaging scheme}
\label{sec:averagingscheme}

We now introduce the averaging scheme used to quantify averaged dynamics in this paper. This averaging formalism is a direct generalization of that presented in \cite{generalbackreac2}, the difference being that we allow for an arbitrary volume measure on the selected hypersurfaces. We discuss the interpretation of the generalized volume measure, and highlight several relevant subcases of the averaging scheme in relation to the existing literature. 

\subsection{The window function}
Following \cite{generalbackreac1,generalbackreac2} we consider scalar functions integrated over space-time domains that are selected out of the space-time $4-$manifold $\CM$ by appropriate choices of window functions. 
In the context of this paper we shall consider window functions that single out compact regions of $3-$dimensional spatial hypersurfaces. 
Averaging over $3-$dimensional hypersurfaces is natural when we want to describe the evolution of averaged properties of spatial sections of the Universe.

Here we shall consider a slightly broader class of $3+1$ window functions than in \cite{generalbackreac1,generalbackreac2}, to allow for arbitrary positive volume measures on the hypersurface of integration. 
Hence, we do not restrict ourselves to having the volume measure coincide with the adapted volume measure in the frame of the foliation. Such a more general volume measure is natural in several settings, some of which we shall investigate below. 
This furthermore allows us to make explicit which properties of the averaged expressions are related to the foliation and which are related to the volume measure. 
When investigating foliation dependence \cite{foliationdependence} the separation of these contributions will be useful. 

We shall consider the broad class of window functions 
\begin{equation}
 \label{eq:window} \fl
\; W_{\Sc, \Sc_0, \Bound, \Bound_0, \VecField}
= - \Vec^{\mu} \nabla_{\mu}( \heavy (\Sc_0 - \Sc) )\heavy (\Bound_0 - \Bound) \, = (\Vec^\mu \nabla_\mu \Sc) \, \deltafun(\Sc_0 - \Sc) \heavy(\Bound_0 - \Bound)\, ,
\end{equation} 
where $\Sc$ is a scalar with time-like gradient that determines the spatial foliation of integration (with hypersurfaces $\Sc = const.$) and $\Bound$ is a scalar with space-like (or possibly null) gradient that is used to bound the averaging domain. $\Sc_0$ and $\Bound_0$ are constants that respectively select a specific hypersurface of the foliation ($\Sc = \Sc_0$) and the domain's spatial boundary ($\Bound = \Bound_0$).
$\VecField$ is an arbitrary time-like vector field, that need not be normalized, and that will in general not be normal to the hypersurfaces defined by $\Sc$. $\heavy$ is the unit step function; we use the convention $\heavy(0) = 1$ throughout.
We shall call $\Sc$ the hypersurface scalar, $\Bound$ the boundary scalar, and $\VecField$ the volume measure vector. We shall drop the subscripts denoting the dependencies of $W$ in the following.

This form of the window function generalizes that of \cite{generalbackreac2} through the freedom of choice of the volume measure vector, which in \cite{generalbackreac2} is restricted to being the unit normal vector $\bm n$ to the hypersurfaces defined by $\Sc$.
$\VecField$ determines the volume measure on the hypersurfaces defined by $\Sc$. This corresponds to considering the usual oriented volume element 
\bea \label{eq:induced_vol} \fl
\qquad  {\mathrm d}V^{\lambda} &= -  n^{\lambda}    \frac{\sqrt{g}}{6} \, n^{\mu} \epsilon_{\mu \nu \varrho \sigma} \, \mathrm{d}x^{\nu} \wedge \mathrm{d}x^{\varrho} \wedge \mathrm{d}x^{\sigma} \qquad ; \qquad    n_{\mu} = \frac{- \nabla_{\mu} \Sc }{( -g^{\nu \sigma} \nabla_{\nu} \Sc \nabla_{\sigma} \Sc)^{1/2} } \ ,  
\eea 
(where $g \equiv - \det \left( g_{\mu \nu} \right)$, and $\epsilon$ is the Levi-Civita symbol) projected along the vector $\VecField$. 
Thus, the integration measure that we use on the surfaces defined by constant $\Sc$ is
\bea \label{eq:induced_volumemeasure}
\mathrm{d} \Vol  \equiv \Vec_{\mu} \, \mathrm{d}V^{\mu} \; .
\eea 
We can think of $\Vec_{\mu} \, {\mathrm d}V^{\mu}$ as the flux of $\VecField$ through the infinitesimal volume ${\mathrm d}V^{\mu}$.

If $\VecField$ is taken to be the normal vector $\bm n$ to the $A=const.$ hypersurfaces, we simply recover the Riemannian volume measure of the hypersurfaces, $\mathrm{d}\Vol = n_\mu \, \mathrm{d} V^\mu$.
Alternatively, we may take the volume measure vector $\VecField$ to be a $4-$velocity field $\bm u$ of physical interest, in general tilted with respect to the normal $\bm n$.
In this case, the integration measure defined in (\ref{eq:induced_volumemeasure}) becomes 
\bea \label{eq:induced_vol_measure}
\mathrm{d} \Vol & \equiv u_{\mu} \, \mathrm{d}V^{\mu} = - u_{\mu} n^{\mu}  \frac{\sqrt{g}}{6} \, n^{\lambda} \epsilon_{\lambda \nu \varrho \sigma} \, \mathrm{d}x^{\nu} \wedge \mathrm{d}x^{\varrho} \wedge \mathrm{d}x^{\sigma}  \nonumber\\
&= \gamma \frac{\sqrt{g}}{6} \, n^{\lambda} \epsilon_{\lambda \nu \varrho \sigma} \, \mathrm{d}x^{\nu} \wedge \mathrm{d}x^{\varrho} \wedge \mathrm{d}x^{\sigma}  
 \nonumber \\
 &= \gamma \frac{\sqrt{g}}{6} \, (- \nabla_{\nu} \Sc   \nabla^{\nu} \Sc  )^{-1/2}  \, \epsilon_{\; ijk} \, \mathrm{d}\bar{x}^i \wedge \mathrm{d}\bar{x}^j \wedge \mathrm{d}\bar{x}^k = \gamma\, n_\mu \, \mathrm{d}V^\mu\;,
\eea 
where $\bar{x}^\mu = (\Sc,\bar{x}^i)$ is an adapted coordinate system to the foliation of $\Sc$, and where $\gamma \equiv - \bm u \cdot \bm n$ is the tilt, or Lorentz factor, between the normal of the hypersurfaces and the $4-$velocity $\bm u$.
The infinitesimal volume element $\mathrm{d}\Vol$ measures the local proper volume (around $\Sc = \Sc_0$) of the fluid element defined by the infinitesimal fluid flow tube that intersects the hypersurface $\{\Sc=\Sc_0\}$ at the points of the time coordinate (in the $\bar x^\mu$ basis) $\Sc = \Sc_0$ and of the spatial coordinates spanning the range $[\bar x^i, \bar x^i + \mathrm{d} \bar x^i]$. The Riemannian volume measure $n_\mu \, \mathrm{d}V^\mu$ of this fluid element as it intersects the hypersurface $\{\Sc=\Sc_0\}$, is its volume measure in the frame defined by $\bm n$, and it is thus Lorentz-contracted with respect to $\mathrm{d}\Vol$. Hence, the choice $\VecField = \bm u$ introduces a local proper volume measure of the fluid as the Riemannian volume measure multiplied by the local Lorentz factor $\gamma$.

\subsection{Averages of scalars}

We define the integral over a scalar $\Scalar$ over the space-time domain $\{ \Sc = \Sc_0, \, \Bound \leq \Bound_0 \}$ singled out by the window function $W$ as follows:
\begin{equation} \label{eq:integrationdef}
I_{W} (\Scalar) \equiv  \int_{\mathcal{M}} \mathrm{d}^4 x \, \sqrt{g}\,  \Scalar \, W \; ,
\end{equation}
and we define the average of a scalar $\Scalar$ as
\begin{equation} \label{eq:average}
\average{\Scalar}_{W}  \equiv  \frac{\int_{\mathcal{M}}\, \mathrm{d}^4 x \, \sqrt{g}\,  \Scalar \, W}{\int_{\mathcal{M}} \,\mathrm{d}^4 x \, \sqrt{g} \, W} = \frac{I_W(\Scalar)}{\Vol}  \; ,
\end{equation}
where $\Vol \equiv I_W(1)$ is the volume of the domain as measured by $\mathrm{d} \Vol$. 
The functional dependencies of $I_W(\Scalar)$ and $\average{\Scalar}_W$ on the variables of $W$ are kept implicit for ease of notation, and we shall also drop the window function index $W$ in what follows. 

\subsection{Examples of window functions}

We now present several possible choices for the window function, adapted to specific descriptions.

\subsubsection{Riemannian averages:}

As discussed above, the choice $\bm \Vec = \bm n$ implies integration with respect to the Riemannian volume element of the hypersurfaces determined by $\Sc$ in the definitions (\ref{eq:integrationdef})--(\ref{eq:average}) for integration and averages. This choice corresponds to the averaging formalisms that are often used in the literature for general foliations, in addition to specific (not always covariantly defined) conditions on the propagation of the domain boundary (see a comprehensive list of such general foliation extensions of \cite{dust,perfectfluid} in the literature comparison investigated in \cite{generalfluid}). This is the choice made in \cite{generalbackreac2}, where the propagation of the domain is in principle kept general, but is specified as following the normal vector, $\bm n \cdot \bm \nabla \Bound = 0$, when derivation of averaged Einstein equations is considered.

\subsubsection{Lagrangian window functions:}
\label{subsubsec:lagrangianWindowDef}
One can also use the integration measure arising from $\VecField = \bm u$, where $\bm u$ is the generator of flow lines of a physical fluid, together with the requirement of a domain propagating along the fluid flow, $\bm u \cdot \bm \nabla \Bound = 0$. We do not at this point specify the time function $\Sc$. 
We call such a choice a \textit{Lagrangian} window function, since the spatial domain is comoving with the fluid, and the volume measure is defined as the proper volume measure of the fluid elements. 

The proper volume element of the fluid (\ref{eq:induced_vol_measure}) and the associated volume and averages as defined by (\ref{eq:average}) are equivalent to those of \cite{foliations}, here derived from a manifestly covariant window function. This  explicitly shows that all results derived from the integration of scalars with this choice of volume element in \cite{foliations} are covariant, as well as the former results of \cite{dust,perfectfluid} obtained with the same volume element in the case of a fluid-orthogonal foliation ($\VecField = \bm u = \bm n$).

\subsubsection{Mass-weighted averages:}
\label{subsubsec:massweighted}
Consider a fluid with $4-$velocity $\bm u$ and with an associated conserved local rest mass current $\bm M$, 
\bea \label{eq:restmass_current}
M^{\mu} = \varrho u^{\mu} \qquad ;   \qquad  \nabla_{\mu}M^{\mu} = 0 \;,
\eea 
where $\varrho$ is the rest mass density. 
We can define a mass-weighted Lagrangian average by choosing $\Vec^\mu = M^\mu$ in (\ref{eq:window}) and $\bm u \cdot \bm \nabla \Bound = 0$. 
This mass-weighted average corresponds to that formulated for irrotational dust in fluid-orthogonal foliations in \cite{massweighted}, but here expressed in the explicitly covariant formalism and extended to arbitrary fluids and foliations.

\subsubsection{Other weighted averages:}
As illustrated by the previous example, the freedom of choice of $\VecField$ allows for any weighting of the averages. One may thus use the window function (\ref{eq:window}) to define, e.g., averages weighted by curvature, or by other functions related to curvature degrees of freedom in the spirit of the `q-average' of Sussman \cite{sussman1,sussman2}\footnote{Note that the `q-average' is constructed for the specific metrics of the Lema\^itre-Tolman-Bondi and Szekeres models by introducing a weighting in the average that is defined from metric degrees of freedom in a particular coordinate system. It is therefore not formulated in a manifestly covariant way. 
However, we may simply extend the definition of the weighting to any other coordinate system, by requiring the weighing to be invariant under the change of coordinates. With such an extension the weighting function is per construction a 
$4-$scalar, and the `q-average' becomes covariant.}, 
writing the corresponding window function under a manifestly covariant form.

\subsubsection{Extensions to light cone averages:}
\label{WindowFunctionsLightcones}
One may choose a boundary scalar with null gradient such that $\{\Bound=\Bound_0\}$ defines the past light cone of a given event, as studied in \cite{lightconeav1} in the case $\VecField  = \bm n$. Integrals and averages are then taken over the spatial region defined by the interior of the light cone at time $\Sc=\Sc_0$.

Because $\VecField$ is not constrained to be the unit normal vector to the $\Sc=const.$ hypersurfaces, the formalism can also be straightforwardly extended to averaging over past light cones by choosing $\Sc$ as the appropriate scalar with light-like gradient and $\VecField$ as a fixed time-like vector, e.g. the 4-velocity $\bm u$ of a fluid source. One might then also replace $\Bound$ by a scalar of time-like gradient; another averaging operator discussed in \cite{lightconeav1} is recovered in this case if $\VecField$ is taken as the normalized gradient of $\Bound$. For either a space-like or a time-like $\bm \nabla \Bound$, such a window function would then select a bounded part of the past light cone of a given event. The variations of integrals or averages with respect to $\Sc_0$ then provide information on \emph{drift} effects as this event changes, while the description of time evolution along a fixed past light cone would instead require an analysis of variations with respect to $\Bound_0$.

\section{The Buchert-Ehlers commutation rule}
\label{Commutation}

We now give a generalization of the commutation rule \cite{buchertehlers}, \cite{dust,perfectfluid,buchertcarfora,ellisbuchert}, and the corresponding manifestly covariant version \cite{generalbackreac2}.
We focus on different possible rewritings of the commutation rule, which can prove useful for interpretation and for compactness of averaged equations. We then apply it to a Lagrangian window function and to the evolution of the fluid rest mass within the integration domain.

\subsection{General formulation}

The essential insight of scalar averaging schemes is that time-derivatives and averaging operations do not commute in general. 
The commutation rule for the integral can be derived by differentiating the expression for $I(\Scalar)$ in the form (\ref{eq:integrationdef}) with respect to $\Sc_0$: 
\bea \label{eq:commutation_rule} 
\fl
\qquad I(\Scalar)' {}= \int_{\mathcal{M}} \, \mathrm{d}^4 x \sqrt{g} \, \Scalar \,  \Vec^{\nu} \nabla_{\nu} \Sc  \left(  \frac{\partial}{\partial \Sc_0} \deltafun(\Sc_0 - \Sc) \right) \heavy(\Bound_0 - \Bound)     \nonumber \\
\fl \qquad \qquad
{} = \int_{\mathcal{M}} \, \mathrm{d}^4 x \sqrt{g} \, \Scalar \,  \Vec^{\nu} \nabla_{\nu} \Sc  \left( - \frac{\partial}{\partial \Sc} \deltafun(\Sc_0 - \Sc) \right) \heavy(\Bound_0 - \Bound)      \nonumber \\
\fl \qquad \qquad  {} = \int_{\mathcal{M}} \, \mathrm{d}^4 x \sqrt{g} \, \Scalar \,  \Vec^{\nu} \nabla_{\nu} \Sc  \left( - \frac{Z^{\mu}}{Z^{\nu}\nabla_{\nu} \Sc} \nabla_{\mu} \deltafun(\Sc_0 - \Sc) \right) \heavy(\Bound_0 - \Bound)     \nonumber \\
\fl \qquad \qquad  {} = \int_{\mathcal{M}} \, \mathrm{d}^4 x \sqrt{g} \,  W    \frac{\nabla_{\mu} \left( \Scalar Z^{\mu} \frac{\Vec^{\kappa} \nabla_{\kappa} \Sc}{ Z^{\sigma} \nabla_{\sigma} \Sc } \, \heavy (\Bound_0 - \Bound)   \right) }{\Vec^{\nu} \nabla_{\nu} \Sc} \nonumber \\
\fl \qquad \qquad {} = I \left(  \frac{Z^{\mu} \nabla_{\mu} \Scalar }{ Z^{\sigma} \nabla_{\sigma} \Sc  }      \right)  +   I \left(  \frac{\Scalar \, \nabla_{\mu}  \left( Z^{\mu} \frac{\Vec^{\kappa} \nabla_{\kappa} \Sc}{ Z^{\sigma} \nabla_{\sigma} \Sc }   \right)   }{\Vec^{\nu} \nabla_{\nu} \Sc }   \right)    -   I \left(  \frac{ \Scalar \, Z^{\mu} \nabla_{\mu} \Bound  \; \deltafun(\Bound_0 - \Bound)   }{ Z^{\sigma} \nabla_{\sigma} \Sc  }     \right)   \;  , 
\eea 
with the notation $' \equiv \partial / \partial \Sc_0$, and where $\bm Z$ is an arbitrary vector field obeying $\bm Z \cdot \bm \nabla \Sc \neq 0$ everywhere. 
The third line of (\ref{eq:commutation_rule}) follows from $\bm Z \cdot \bm \nabla ( \deltafun(\Sc_0 - \Sc) ) = (\bm Z \cdot \bm \nabla \Sc) \,  \partial_{\Sc} (\deltafun(\Sc_0 - \Sc))$, and the fourth line follows from partial integration, with the convention $\heavy(0) = 1$ implying $\heavy(x) \deltafun(x) = \deltafun(x)$.

$\bm Z$ represents the freedom of the direction in which we define local time derivatives with respect to $\Sc$. 
Non-commutativity is given by the failure of the boundary to be parallel-transported along $\bm Z / (\bm Z \cdot \bm \nabla \Sc)$ and by the change of volume measure along the flow lines of $\bm Z / (\bm Z \cdot \bm \nabla \Sc)$. 
We denote the first term of (\ref{eq:commutation_rule}) the evolution term, the second term the expansion term, and the third term the boundary term.  

The full result (\ref{eq:commutation_rule}) is not dependent on $\bm Z$, but different choices of $\bm Z$ allow us to trade between the three terms in (\ref{eq:commutation_rule}). For instance,
we can make the boundary terms disappear by choosing $\bm Z$ such that $\bm Z \cdot \bm \nabla \Bound = 0$,\footnotemark
\footnotetext{Taking $\bm Z$ to be time-like or null automatically ensures $\bm Z \cdot \bm \nabla \Sc \neq 0$ if $\bm \nabla \Sc$ is time-like.}
i.e., the boundary term contribution does not appear if the direction chosen for time derivation follows the propagation of the boundary. 
Similarly, we might make the evolution term vanish by choosing a $\bm Z$ such that $\bm Z \cdot \bm \nabla \Scalar = 0$. \footnotemark
\footnotetext{Note, however, that if $ \bm \nabla \Scalar \propto \bm \nabla \Sc$, then this choice is not possible, and the evolution term cannot be put to zero.}
The rate of evolution of the volume $I(1)$ and the commutation rule for the average follow from (\ref{eq:commutation_rule}) and are given respectively by
\bea
\label{eq:volume_evol}
\frac{I(1)'}{I(1)} = \average{\frac{\nabla_{\mu}  \left( Z^{\mu} \frac{\Vec^{\kappa} \nabla_{\kappa} \Sc}{ Z^{\sigma} \nabla_{\sigma} \Sc }   \right)   }{\Vec^{\nu} \nabla_{\nu} \Sc } } - \average{\frac{Z^{\mu} \nabla_{\mu} \Bound  \; \deltafun(\Bound_0 - \Bound)   }{ Z^{\sigma} \nabla_{\sigma} \Sc  }} \; ; 
\\
\label{eq:commutation_rule_average}
\average{\Scalar}' = \frac{I (\Scalar)' }{I (1)} - \average{\Scalar}  \frac{   I(1)' }{I (1)}  = \nonumber\\
\fl \;
  \average{  \frac{Z^{\mu} \nabla_{\mu} \Scalar }{ Z^{\sigma} \nabla_{\sigma} \Sc  }   }  +   \average{  \frac{\big( \Scalar - \average{\Scalar }  \! \big) \,  \nabla_{\mu}  \left( Z^{\mu} \frac{\Vec^{\kappa} \nabla_{\kappa} \Sc}{ Z^{\sigma} \nabla_{\sigma} \Sc }   \right)   }{\Vec^{\nu} \nabla_{\nu} \Sc }  }    - \average{   \frac{ \big( \Scalar - \average{\Scalar} \!\big)  \, Z^{\mu} \nabla_{\mu} \Bound  \; \deltafun(\Bound_0 - \Bound)   }{ Z^{\sigma} \nabla_{\sigma} \Sc  }    }      .
\eea 
Again, we might trade between the three terms in (\ref{eq:commutation_rule_average}) by changing $\bm Z$, e.g., we can still make the third term vanish by choosing $\bm Z$ to be a time-like vector field comoving with the spatial boundaries of the domain. 

When it is possible to choose a time-like $\bm Z$ such that $ \nabla_{\mu}  \left( Z^{\mu} \frac{\Vec^{\kappa} \nabla_{\kappa} \Sc}{ Z^{\sigma} \nabla_{\sigma} \Sc }   \right) = 0$, and $Z^{\mu} \nabla_{\mu} \Bound = 0$ simultaneously, there is a sense in which time-derivative and the averaging operation commute in (\ref{eq:commutation_rule}) and (\ref{eq:commutation_rule_average}): in this case it is possible to construct flow lines along which the only contribution to the change of $\average{\Scalar}$ is the change of $\Scalar$ itself.
This is the case for a mass-weighted window function (see section \ref{subsubsec:massweighted}). In this case, $\bm Z = \bm u$ satisfies the above requirements, so that the commutation rule (\ref{eq:commutation_rule_average}) reduces to 
\bea \label{eq:commutation_rule_average_Mass}
&\average{\Scalar}' =  \average{  \frac{u^{\mu} \nabla_{\mu} \Scalar }{ u^{\sigma} \nabla_{\sigma} \Sc  }   }  . 
\eea 
Hence, there is commutation of this particular averaging operation and time-derivative along the flow lines of $\bm u$, generalizing this result obtained for irrotational dust in the fluid-orthogonal foliation \cite{massweighted}. This commutation is, however, obtained at the expense of a more complicated definition required for a physical volume (and associated scale factor).
In this setting, the \say{volume} $I(1)$ actually corresponds to a total rest mass within the integration domain, as described in section \ref{massdef}. Thus, as noticed in \cite{massweighted}, defining a physical volume would require to compensate for the weighting by $\varrho$, e.g. by considering $I(1/\varrho)$.

We may choose $\bm Z$ to be the most convenient vector field for simplifying the commutation rules, or may choose it from a geometric motivation as, e.g. in \cite{generalbackreac2}, where $\bm Z$ is chosen to coincide with the normal to the hypersurfaces. 
Alternatively, one may choose a physical vector field for $\bm Z$, e.g. $\bm Z = \bm u$, where $\bm u$ is the $4-$velocity of a physical fluid of interest. In this formulation the terms in (\ref{eq:commutation_rule}) and (\ref{eq:commutation_rule_average}) can be interpreted in terms of evolution along physical flow lines of a fluid and its expansion. 

\subsection{Application to the case of a Lagrangian window function}
\label{lagrangianWindowComm}

Let us consider a Lagrangian window function as defined in section \ref{subsubsec:lagrangianWindowDef}. Writing the commutation rule (\ref{eq:commutation_rule}) with $\bm Z = \bm u$ we have in this case
\bea \label{eq:commutation_rule_Lagrangian}
& I(\Scalar)' = I \left(  \frac{u^{\mu} \nabla_{\mu} \Scalar }{ u^{\sigma} \nabla_{\sigma} \Sc  }      \right)  +   I \left(  \frac{\Scalar \,  \nabla_{\mu} u^{\mu}   }{u^{\sigma} \nabla_{\sigma} \Sc }   \right)   \quad ; \quad I(1)' = I \left(\frac{\nabla_\mu u^\mu}{u^\sigma \nabla_\sigma \Sc} \right)  \; ,
\eea 
where the first contribution comes from the change of $\Scalar$ along the flow lines of $\bm u$, and the second contribution from the expansion $\nabla_\mu u^\mu$ of the fluid. 
Note the normalization $u^{\sigma} \nabla_{\sigma} \Sc$, which is a change of measure between the proper time parameter $\tau$ of the fluid and the 
foliation parameter $\Sc$ along each fluid flow line. Hence, this normalization reduces to unity if and only if $\Sc$ is a proper time of $\bm u$.
 
The analogous commutation rule for the average (\ref{eq:commutation_rule_average}) yields
\bea \label{eq:commutation_rule_average_Lagrangian}
 &\average{\Scalar}' = \average{  \frac{u^{\mu} \nabla_{\mu} \Scalar }{ u^{\sigma} \nabla_{\sigma} \Sc  }   }  +   \average{  \frac{( \Scalar - \average{\Scalar }   )   \nabla_{\mu}  u^{\mu}    }{u^{\sigma} \nabla_{\sigma} \Sc }  }   .
\eea 
There are at least two natural ways of choosing $\Sc$ in the Lagrangian spirit of formulating the window function.
In cases where $\bm u$ is irrotational, it is then proportional to the gradient of a scalar $\alpha$, and we can choose $\Sc$ to define a foliation in the rest frame of the fluid (i.e. fluid-orthogonal hypersurfaces) by $\Sc = \alpha$. 
An alternative natural choice of $\Sc$ is a proper time parameter $\tau$ of $\bm u$ \cite{foliations,generalfluid}. This has the advantage of being always possible, even if $\bm u$ has vorticity, and of providing a clear physical interpretation of $\Sc$ as the time parameter in evolution equations for average quantities.  However, the time-like nature of $\bm \nabla \tau$ can in general not be guaranteed. Note that the above conditions define classes of foliation scalars, i.e. further specifications are required to determine them uniquely.\footnote{%
The \textit{proper time foliation} $\Sc = \tau$ is only specified up to an additive function $\beta$ obeying $\bm u \cdot \bm \nabla \beta = 0$. 
The \textit{fluid frame foliation} $\Sc = \alpha$ is only specified up to a reparametrization, $\Sc = f(\alpha)$, for any non-decreasing function $f$ of $\alpha$. This freedom can be denoted a gauge freedom, since it can be viewed as a time 
reparametrization within the original foliation itself. See \ref{gauge1} for further details on gauge freedom in the labeling of hypersurfaces.
}
A choice of proper time foliation can be simultaneously fluid-orthogonal only when the fluid is irrotational and 
geodesic.\footnote{A fluid-orthogonal foliation implies that $\bm u = \bm n = - N \bm \nabla \Sc$ with the lapse $N = (- \bm \nabla \Sc \cdot \bm \nabla \Sc)^{-1/2}$.
The vorticity of $\bm u$ thus has to vanish, which is part of Frobenius' theorem. It also implies that the $4-$acceleration $\bm a$ of the fluid relates to the lapse variations as $a^\mu = N^{-1} \, b^{\mu \nu} \nabla_\nu N$ \cite{gourg:foliation,foliations},
with $\bm b$ the fluid-orthogonal projector. If $\Sc$ is additionally required to be a proper time function for the fluid,  $\bm u \cdot \bm \nabla \Sc = 1$, then $N=1$ everywhere and $\bm a = 0$.
This shows that the fluid flow must also be geodesic.}

\subsection{Total rest mass of the averaging domain}
\label{massdef}

Consider a conserved local rest mass current $M^{\mu} = \rho u^{\mu}$ as in (\ref{eq:restmass_current}).
We can define a total rest mass within the domain at $\Sc = \Sc_0$ as
\begin{equation} \label{eq:massdef}
M(\Sc_0) \equiv  \int_{\mathcal{M}} \mathrm{d}^4 x \, \sqrt{g} \,  M^{\mu}\nabla_{\mu} ( \heavy (\Sc - \Sc_0))  \heavy (\Bound_0 - \Bound)    \; ,        
\end{equation}
i.e., as $I(1)$ for a window function with $\Vec^\mu = M^\mu$ (e.g. the mass-weighted window function, see section \ref{subsubsec:massweighted}). Applying (\ref{eq:commutation_rule}) gives the evolution of $M(\Sc_0)$ which, due to the local conservation of $M^{\mu}$, reduces to a single boundary term
\begin{equation} \label{eq:massevolution}
M(\Sc_0)'  = - \int_{\mathcal{M}} \mathrm{d}^4 x  \,\sqrt{g} \, M^{\mu} \nabla_{\mu} \Bound  \; \heavy (\Sc - \Sc_0) \, \deltafun (\Bound_0 - \Bound) \; , 
\end{equation}
i.e. the evolution of mass is given by the flux of the mass current $M^{\mu}$ out of the averaging domain. Thus, $M(\Sc_0)$ is constant in $\Sc_0$ when the domain is comoving with the fluid elements, $\bm u \cdot \bm \nabla \Bound = 0$.
For such a comoving integration domain, $M = M(\Sc_0)$ (for any $\Sc_0$), as defined by (\ref{eq:massdef}), corresponds to the total conserved rest mass of the fluid within the domain. In this case, the additional requirement $\VecField = \bm u$ sets a Lagrangian window function (as defined in section \ref{subsubsec:lagrangianWindowDef}). The conserved total rest mass within the domain then takes the natural form $M = I(\varrho)$. For other volume measures, in general, $I(\varrho)$ would not correspond to the rest mass within the domain and would not be conserved, due to a weighting or due to the volume not being measured in the fluid's local rest frames. (For instance, for the hypersurfaces Riemannian volume measure, $\VecField = \bm n$, and still for a comoving domain, the integrated rest mass would have to be written $M = I(\gamma \varrho)$ with $\gamma = - \bm n \cdot \bm u$.) A Lagrangian window function $\{\VecField = \bm u$, $\bm u \cdot \bm \nabla \Bound = 0 \}$ thus appears as a particularly natural choice to follow and characterize a given collection of fluid elements, if a preferred fluid frame with an associated rest mass current is present in the model universe.
We shall focus again in section \ref{averaged_comoving} on domains that follow the  propagation of the fluid---hence preserving the associated rest mass---as a subcase of particular interest of more general averaged evolution equations, to which we turn now.

\section{The averaged Einstein equations}
\label{sec:AEE}

The general averaging formalism and the commutation rule are applied below to scalar projections of the Einstein equations. The resulting system of averaged evolution equations allows for a covariant definition of \emph{cosmological backreaction terms}. We shall then explicitly provide the simpler form taken by these equations for a domain that follows the fluid world lines, and we discuss the natural choices $\VecField = \bm n$ and $\VecField = \bm u$.

\subsection{Local variables and relations}

In this subsection we consider an averaging domain defined by a time-like propagation of its boundary. We thus assume that a unit time-like propagation vector field $\bm P$ can be defined such that it satisfies $\bm P \cdot \bm \nabla \Bound = 0$, at least on the domain's boundary $\{\Bound = \Bound_0 \}$. Applying the commutation rules (\ref{eq:commutation_rule})--(\ref{eq:commutation_rule_average}) with the choice $\bm Z = \bm P$ will then ensure the vanishing of the boundary terms in these equations.

Kinematic variables may then be defined for this vector field by decomposing its gradient with respect to $\bm P$ and its null-space as follows, using the orthogonal projector $\mathbf k$ with components $k_{\mu \nu} = g_{\mu \nu} + P_\mu P_\nu$:
\bea \label{eq:expansionP}\fl
 \quad  \nabla_{\mu} P_{\nu}  = - P_\mu a^P_\nu + \frac{1}{3} \Theta_P \, k_{\mu \nu} + \sigma^P_{\mu \nu} + \omega^P_{\mu \nu} \, ; \nonumber \\
 \fl
  \quad a^P_\mu = P^\nu \nabla_\nu P_\mu \,;\; \Theta_P = k^{\mu \nu} \nabla_\mu P_\nu  \,;\;    \sigma^P_{\mu \nu} = k^\alpha_{\,(\mu} k^\beta_{\,\nu)} \nabla_\alpha P_\beta - \frac{1}{3} \Theta_P\, k_{\mu \nu} \,;\;  \omega^P_{\mu \nu} =  k^\alpha_{\,[\mu} k^\beta_{\,\nu]} \nabla_\alpha P_\beta \,; \nonumber \\
 \fl
 \quad \sigma_P^2 = \frac{1}{2} \, \sigma^P_{\mu \nu} \,\sigma^{P, \mu \nu} \;;  \ \  \omega_P^2 = \frac{1}{2} \, \omega^P_{\mu \nu} \,\omega^{P, \mu \nu}\, .  
\eea
Assuming the presence of a preferred non-singular fluid flow as a source, with $4-$velocity $\bm u$, the (fully general) energy-momentum tensor is naturally decomposed with respect to $\bm u$ and its null-space:
\bea
\label{eq:defEnergy}
\fl
\qquad   T_{\mu \nu} = \epsilon \, u_{\mu}  u_{\nu} + 2 \,q_{( \mu} u_{\nu )}  + p \, b_{\mu \nu} + \pi_{\mu \nu} \;; \nonumber\\
\fl \qquad  \epsilon  \equiv u^{\mu} u^{\nu}T_{\mu \nu} \;;  \quad  q_{\mu} \equiv - b^{\alpha}_{\; \mu}  u^{\beta} T_{\alpha \beta} \;;  \quad    p  \equiv \frac{1}{3} \, b^{\mu \nu}T_{\mu \nu} \;;\quad  \pi_{\mu \nu} \equiv  b^{\alpha}_{\; \mu} b^{\beta}_{\; \nu} T_{\alpha \beta} - p\, b_{\mu \nu} \; ,   
\eea 
where $\mathbf{b}$ is the projector onto the fluid's rest frames, with components $b_{\mu \nu} = g_{\mu \nu} + u_\mu u_\nu$.
It may alternatively be decomposed using $\bm P$. In particular, one can define the energy density $E_P$ and pressure $S_P/3$, in the frames defined by $\bm P$, from, respectively:
\bea
E_P \equiv P^\mu P^\nu T_{\mu \nu} \; ; \quad S_P = k^{\mu \nu} T_{\mu \nu} \; .
\eea
These variables are related to the fluid rest frame energy density $\epsilon$, pressure $p$, and to the non-perfect fluid contributions \textit{via}
\begin{equation} \fl
\label{eq:energy_difference} 
E_P - \epsilon = \frac{1}{2} \big[E_P + S_P - (\epsilon + 3 p) \big] = (\epsilon + p) \! \left[ (u^\mu P_\mu)^2 - 1 \right] + 2 \,(u^\mu P_\mu) (P^\nu q_\nu) + \pi_{\mu \nu} P^\mu P^\nu \, . \;
\end{equation}
The following Raychaudhuri equation for $\bm P$ is then obtained by combining the Einstein equation projected twice along $\bm P$, and its trace:
\bea
\label{eq:RaychaudhuriP}
P^\mu \nabla_\mu \Theta_P = - \frac{1}{3} \Theta_P^2 - 2 \sigma_P^2 + 2 \omega_P^2 + \nabla^\mu a^P_\mu - 4 \pi G (E_P + S_P) + \Lambda \; .
\eea
We define an effective scalar $3-$curvature for the null-space of $\bm P$ (which is not hypersurface-forming if $\omega_P^2 \neq 0$) as follows:
\bea \label{eq:3curvP}
\mathcal{R}_P &\equiv \nabla_{\mu}P^{\nu} \, \nabla_{\nu}P^{\mu} - \nabla_{\mu}P^{\mu} \, \nabla_{\nu}P^{\nu} + R + 2R_{\mu \nu} P^{\mu} P^{\nu} \; .
\eea
This definition of effective $3-$curvature reduces to the scalar $3-$curvature of the $\bm P$-orthogonal hypersurfaces when they exist (i.e., for $\omega_P^2 = 0$, by Frobenius' theorem). Such a generalization of the hypersurface-based notion is not unique; we here follow a similar definition as that of, e.g. \cite{ellis:vorticity}. This convention implies the following relation in the form of an energy constraint:
\bea \label{eq:HamiltonianP}
& \frac{2}{3} \Theta_P^2 = - \mathcal{R}_P + 2 \sigma_P^2 - 2 \omega_P^2  + 16 \pi G \, E_P + 2 \Lambda\;.
\eea

\subsection{Averaged evolution equations}
\label{subsecAveragedEqsGeneral}

We use the general window function (\ref{eq:window}) and define an effective \say{scale factor} $a$ as $a = (I(1)/I(1)_\mathbf{i})^{1/3}$, where the subscript $\mathbf{i}$ denotes a value on some initial hypersurface $\Sc=\Sc_\mathbf{i}$.

As noted for the example of the mass-weighted average \cite{massweighted}, it should be kept in mind that this definition is only relevant as a scale factor if it can be interpreted as a typical length derived from a volume, i.e. only when the choice of integration measure defined by $\VecField$ allows for the interpretation of $I(1)$ as a volume. Another definition of `scale factor' that does relate it to a physical volume (e.g. to $I(1/\varrho)$ in the case of the mass-weighted average) may otherwise be more appropriate.
It should also be noted, that the effective \say{scale factor} $a$ in general does not have an interpretation in terms of mean redshift of null bundles (the averaging scheme presented in this paper is too general to make a direct link to statistical light propagation). However, when $I(1)$ does measure a volume, and under the assumptions that (i) the frame of averaging is associated with statistical homogeneity and isotropy, that (ii) structures are slowly evolving (allowing null-rays to probe the statistical homogeneity scale), and that (iii) typical emitters and observers of light are reasonably close to being in the averaging frame, $a$ might be interpreted as the inverse of a \say{statistical redshift} averaged over many observers and emitters \cite{rasanen}. More generally, only assuming a choice of window function such that $I(1)$ measures a physical volume, $a$ should merely be interpreted as an effective length scale of an averaging region defined in a given foliation.

Averaging the above equations (\ref{eq:HamiltonianP}) and (\ref{eq:3curvP}) with the averaging definition (\ref{eq:average}), and making use of the volume evolution rate (\ref{eq:volume_evol}) and the commutation rule (\ref{eq:commutation_rule_average}) with the choice $\bm Z = \bm P$, implying $\bm Z \cdot \bm \nabla \Bound = 0$, yields the following evolution equations for $a$:
\bea
\label{eq:HamiltonianAveragedP}
\fl  3 \left( \frac{a'}{a} \right)^2 \! &= 8 \pi G \average{\frac{\epsilon}{(P^\mu \nabla_\mu \Sc)^2}} + \Lambda \average{\frac{1}{(P^\mu \nabla_\mu \Sc)^2}} - \frac{1}{2} \average{\frac{\mathcal{R}_P}{(P^\mu \nabla_\mu \Sc)^2}} - \frac{1}{2} \CQ - \frac{1}{2} \CT \, ; \\
\label{eq:RaychaudhuriAveragedP}
\fl \quad \; \; 3 \frac{a''}{a} &= - 4 \pi G \average{\frac{\epsilon + 3 p}{(P^\mu \nabla_\mu \Sc)^2}} + \Lambda \average{\frac{1}{(P^\mu \nabla_\mu \Sc)^2}} + \CQ + \CP + \frac{1}{2} \CT \, .
\eea
These equations feature three backreaction terms, a \emph{kinematical backreaction} $\CQ$, a \emph{dynamical backreaction} $\CP$, and an \emph{energy-momentum backreaction} $\CT$ that captures the difference of  the energy densities as measured in two different frames (see \cite{generalfluid}). These backreaction 
terms are defined as follows:
\begin{eqnarray}
\fl \CQ \equiv \frac{2}{3} \left[\average{\frac{\Theta^2_P}{(P^\rho \nabla_\rho \Sc)^2}} - \average{\frac{\Theta_P + \Ratio^{-1} \, P^\mu \nabla_\mu \Ratio}{P^\rho \nabla_\rho \Sc}}^{\!2}\right] -  \average{\frac{2\sigma_P^2}{(P^\mu \nabla_\mu \Sc)^2}} \! + \average{\frac{2\omega_P^2}{(P^\mu \nabla_\mu \Sc)^2}}; \nonumber \\
\CP \equiv \average{\frac{\nabla^\mu a_\mu^P}{(P^\mu \nabla_\mu \Sc)^2}} + \average{\frac{\Theta_P}{(P^\rho \nabla_\rho \Sc)^2} \left( 2 \, \frac{P^\mu \nabla_\mu \Ratio}{\Ratio} - \frac{P^\mu \nabla_\mu (P^\nu \nabla_\nu \Sc)}{P^\sigma \nabla_\sigma \Sc} \right) } \nonumber \\
{} + \average{\frac{\Ratio^{-1} \, P^\mu \nabla_\mu (P^\nu \nabla_\nu \Ratio)}{(P^\mu \nabla_\mu \Sc)^2}} - \average{\frac{\Ratio^{-1} \, P^\mu \nabla_\mu \Ratio}{(P^\rho \nabla_\rho \Sc)^2} \; \frac{P^\nu \nabla_\nu (P^\kappa \nabla_\kappa \Sc)}{P^\sigma \nabla_\sigma \Sc}} \; ; \nonumber\\
 \CT = -16 \pi G \average{\frac{E_P - \epsilon}{(P^\mu \nabla_\mu \Sc)^2}} \; ,
\end{eqnarray}
with the energy difference $E_P - \epsilon$ given by (\ref{eq:energy_difference}), and with the ratio of \say{Lorentz factors} $\Ratio \equiv (\Vec^\mu \nabla_\mu \Sc) / (P^\nu \nabla_\nu \Sc) = (-\Vec^\mu n_\mu) / (- P^\nu n_\nu)$, $-\Vec^\mu n_\mu$ being a Lorentz factor when $\VecField$ is normalized.

From the requirement of (\ref{eq:HamiltonianAveragedP}) being the integral of (\ref{eq:RaychaudhuriAveragedP}) we get the integrability condition:
\bea \label{eq:IntegrabilityP} \fl
\qquad  \CQ'  + 6   \frac{ a'}{a} \CQ  +  \average{  \frac{  \CR_P    }{ (P^{\sigma} \nabla_{\sigma} \Sc  )^2}   }'   +  2  \frac{a' }{a}  \average{  \frac{  \CR_P }{ (P^{\sigma} \nabla_{\sigma} \Sc  )^2}   }   + \CT' + 4 \frac{a'}{a} \CT + 4   \frac{ a'}{a} \CP    \nonumber   \\
\fl \qquad \qquad = 16 \pi G \left(     \average{  \frac{  \epsilon   }{ (P^{\sigma} \nabla_{\sigma} \Sc  )^2}   }'  + 3  \frac{ a' }{a}     \average{  \frac{  \epsilon + p   }{ (P^{\sigma} \nabla_{\sigma} \Sc  )^2}   }      \right) + 2 \Lambda \average{(P^\sigma \nabla_\sigma \Sc)^{-2} }' .   
\eea
Defining the kinematic variables of the fluid from the decomposition of the $4-$velocity gradient,
\bea
\label{eq:expansion}
\fl
 \qquad  \nabla_{\mu} u_{\nu}  = - u_\mu a_\nu + \frac{1}{3} \Theta \, b_{\mu \nu} + \sigma_{\mu \nu} + \omega_{\mu \nu} \, ; \nonumber \\
 \fl
  \qquad a_\mu = u^\nu \nabla_\nu u_\mu \,;\; \Theta = b^{\mu \nu} \nabla_\mu u_\nu  \,;\;    \sigma_{\mu \nu} = b^\alpha_{\,(\mu} b^\beta_{\,\nu)} \nabla_\alpha u_\beta - \frac{1}{3} \Theta\, b_{\mu \nu} \,;\;  \omega_{\mu \nu} =  b^\alpha_{\,[\mu} b^\beta_{\,\nu]} \nabla_\alpha u_\beta \,; \nonumber \\
 \fl
 \qquad \sigma^2 = \frac{1}{2}     \sigma_{\mu \nu} \sigma^{\mu \nu} \;;  \ \  \omega^2 = \frac{1}{2} \omega_{\mu \nu} \omega^{\mu \nu}\, ,  
\eea
we can express the energy-momentum conservation equation projected onto the fluid frame as follows:
\bea
\label{eq:EnergyCons}
& - u^{\mu} \nabla_{\nu} T^{\nu}_{\; \mu} =u^{\mu} \nabla_{\mu} \epsilon + \Theta (\epsilon + p) + a^{\mu}q_{\mu} + \nabla_{\mu}q^{\mu} + \pi_{\mu \nu} \, \sigma^{\mu \nu}  = 0 \;.
\eea
One can then divide this relation by $(P^\mu \nabla_\mu \Sc)^2$, take the average and apply the commutation rule (\ref{eq:commutation_rule}) with $\bm Z = \bm u$. This yields the average energy conservation law satisfied by the right-hand side of (\ref{eq:IntegrabilityP}):
\bea
\fl \average{\frac{\epsilon}{(P^\sigma \nabla_\sigma \Sc)^2}}' + 3 \frac{a'}{a} \average{\frac{\epsilon+p}{(P^\sigma \nabla_\sigma \Sc)^2}}  = - \average{\frac{\Theta}{\dot \Sc} \, \frac{p }{(P^\sigma \nabla_\sigma \Sc)^2}} + \average{\frac{\Theta}{\dot \Sc}} \average{\frac{p}{(P^\sigma \nabla_\sigma \Sc)^2}} \nonumber \\
\fl {\;} +  \average{\frac{\dot \RatioU / \RatioU}{\dot \Sc} - \frac{(u^\mu \nabla_\mu B) \, \deltafun(\Bound_0 - \Bound)}{\dot \Sc} \!} \average{\frac{p}{(P^\sigma \nabla_\sigma \Sc)^2}} - \average{\! \frac{\epsilon}{(P^\sigma \nabla_\sigma \Sc)^2} \, \frac{(u^\mu \nabla_\mu \Bound) \, \deltafun(\Bound_0 - \Bound)}{\dot \Sc}} \nonumber \\
\label{eq:en_cons_averagedP}
\fl {\;} + \average{\frac{\epsilon}{(P^\sigma \nabla_\sigma \Sc)^2} \, \frac{2 (\dot \Ratio / \Ratio) - (\dot \RatioU / \RatioU) - 2 (\ddot \Sc / \dot \Sc)}{\dot \Sc} } - \average{\frac{a_\mu q^\mu + \nabla_\mu q^\mu + \pi_{\mu \nu} \sigma^{\mu \nu} }{\dot \Sc \, (P^\sigma \nabla_\sigma \Sc)^2 }} \; ,
\eea
with $\RatioU \equiv (\Vec^\mu \nabla_\mu \Sc) / (u^\nu \nabla_\nu \Sc) = (-\Vec^\mu n_\mu) / \gamma$, and using the shorthand notation $\dot \Scalar$ for the proper-time covariant derivative along $\bm u$ of a scalar $\Scalar$, $\dot \Scalar \equiv u^\mu \nabla_\mu \Scalar$. This average conservation equation features two boundary terms that provide the variations in volume and average energy density due to the flux of fluid elements across the domain's boundary if $u^\mu \nabla_\mu \Bound \neq 0$.

The above system of averaged equations (\ref{eq:HamiltonianAveragedP},\ref{eq:RaychaudhuriAveragedP},\ref{eq:IntegrabilityP},\ref{eq:en_cons_averagedP}) is covariant since it only features explicitly covariant terms. The form of these equations is moreover globally preserved under a change of the parametrization of the foliation (using a non-decreasing function of $\Sc$ instead of $\Sc$, preserving the set of hypersurfaces), but the individual terms they contain are not. This is no different from the time-parameter dependence of the expansion and acceleration terms of the Friedmann equations in homogeneous and isotropic cosmologies. This freedom of relabeling the hypersurfaces is important to keep in mind when interpreting averaged evolution equations: as for any parametric equations, e.g. acceleration terms (as second derivatives with respect to a parameter) can be tuned in any desirable way, including the change of sign, by an appropriate change of the parameter. This is discussed in more detail in the specific context of the above averaged equations in \ref{gauge1}. This interpretation issue is simply solved by the choice of a time label with a clear physical meaning for the hypersurfaces. Such a choice can be made specifically for the physical model considered, or from more general conditions, such as taking $\tau$ itself as the parameter $\Sc$ when working within a foliation at constant fluid proper time $\tau$ (see the related remarks that conclude section \ref{lagrangianWindowComm}).

This general set of averaged equations is naturally expressed in terms of geometric variables such as the extrinsic curvature or the intrinsic scalar $3-$curvature of the $A=const.$ hypersurfaces for a domain propagation along the normal vector field, i.e., for $\bm P = \bm n$. In this case, and for $\VecField = \bm n$ (i.e. for Riemannian averages), this system corresponds to the averaged system derived in \cite{generalbackreac2}, with the addition of the integrability condition and the general form of the averaged energy conservation law. 

For a general propagation vector $\bm P$, the explicit contribution of the geometric variables in the above equations can also be recovered by an alternative writing. It can be done by splitting $\bm P$ into a component along $\bm n$ and a component orthogonal to $\bm n$, $\bm P = \gamma_P (\bm n + \bm{v}_P)$ with $\gamma_P = - \bm P \cdot \bm n$ and $\bm n \cdot \bm{v}_P = 0$. The contributions from the decomposition of the gradient of $\bm P$ to the averaged equations can then be expressed in terms of the extrinsic curvature of the hypersurface, e.g. by applying the following split in the commutation rule:
\bea
\fl \frac{ \nabla_{\mu}  \left( P^{\mu} \frac{\Vec^{\rho} \nabla_{\rho} \Sc}{ P^{\sigma} \nabla_{\sigma} \Sc }   \right)   }{\Vec^{\nu} \nabla_{\nu} \Sc } = \frac{\Theta_P + \Ratio^{-1} \, P^\mu \nabla_\mu \Ratio}{P^\rho \nabla_\rho \Sc} = - N \CK + N \frac{\nabla_\mu (\Vec^\nu n_\nu \, v_P^\mu)}{\Vec^\rho n_\rho} + \frac{N n^\mu \nabla_\mu (\Vec^\nu n_\nu)}{\Vec^\rho n_\rho} \; ,\nonumber
\eea
with the lapse function $N \equiv (\nabla^\mu \Sc \nabla_\mu \Sc)^{-1/2}$ and the trace of the extrinsic curvature $\CK \equiv - \nabla_\mu n^\mu$.
The set of equations using this decomposition will then simplify when using the Riemannian volume measure of the hypersurfaces, $\VecField = \bm n$. In the comoving domain case, $\bm P = \bm u$, this returns one of the sets of equations obtained in \cite{generalfluid} when geometric variables--based expressions for the spatial Riemannian volume measure and a domain comoving with the fluid flow are considered.

\subsection{Examples of applications}

\subsubsection{Comoving domains:}
\label{averaged_comoving} 

We now specify the above results to the case of a domain comoving with the fluid, i.e. for which $\bm u \cdot \bm \nabla \Bound = 0$. One can thus take $\bm P = \bm u$. The adapted local Raychaudhuri equation (\ref{eq:RaychaudhuriP}) and energy constraint (\ref{eq:HamiltonianP}) are then expressed in terms of rest frame variables of the fluid:
\bea
 \dot \Theta = - \frac{1}{3} \Theta^2 - 2 \sigma^2 + 2 \omega^2 +  \nabla_\mu a^\mu - 4 \pi G (\epsilon + 3p) + \Lambda \; ; \\
 \frac{2}{3} \Theta^2 = - \CR + 2 \sigma^2 - 2 \omega^2 + 16 \pi G \epsilon + 2 \Lambda \; ,
\eea
with the effective scalar $3-$curvature of the rest frames of $\bm u$ \cite{ellis:vorticity},
\bea
\CR \equiv \nabla_\mu u^\nu \nabla_\nu u^\mu - \nabla_\mu u^\mu \nabla_\nu u^\nu + R + 2 R_{\mu \nu} u^\mu u^\nu \; .
\eea
The corresponding evolution equations for the effective \say{scale factor} $a$ (which may still not be the most appropriate definition in cases where $I(1)$ is not interpreted as a volume) are then written as follows:
\bea
\label{eq:HamiltonianAveraged}
3 \left( \frac{a'}{a} \right)^2 &= 8 \pi G \average{\frac{\epsilon}{{\dot \Sc \strut}^2}} + \Lambda \average{\frac{1}{{\dot \Sc \strut}^{2}}} - \frac{1}{2} \average{\frac{\CR}{ {\dot \Sc \strut}^2}} - \frac{1}{2} \CQ \; ; \\
\label{eq:RaychaudhuriAveraged}
\quad \; 3 \, \frac{a''}{a} &= - 4 \pi G \average{\frac{\epsilon+3p}{{\dot \Sc \strut}^2}} + \Lambda \average{\frac{1}{{\dot \Sc \strut}^2}} + \CQ + \CP \; .
\eea
The energy-momentum backreaction vanishes since $\bm P =\bm u$, and the kinematical and dynamical backreaction terms reduce to the following:
\bea
\fl \quad \CQ \equiv \frac{2}{3} \left( \average{\frac{\Theta^2}{{\dot \Sc \strut}^2}} - \average{\frac{\Theta + \dot \RatioU / \RatioU}{{\dot \Sc \strut}^2}}^2 \right) -2 \average{ \frac{\sigma^2}{{\dot \Sc \strut}^2} } + 2 \average{ \frac{\omega^2}{{\dot \Sc \strut}^2} } \; ; \\
\fl \quad \CP \equiv \average{ \frac{\nabla_\mu a^\mu}{{\dot \Sc \strut}^2} } + \average{\frac{\Theta}{{\dot \Sc \strut}^2} \left( 2 \, \frac{\dot \RatioU}{\RatioU \strut} - \frac{\ddot \Sc}{\dot \Sc \strut} \right) } + \average{\frac{\ddot \RatioU / \RatioU}{{\dot \Sc \strut}^2}} -\average{ \frac{\big(\ddot \Sc / \dot \Sc \big) \, \big(\dot \RatioU / \RatioU \big)}{{\dot \Sc \strut}^2} } \; .
\eea
The integrability condition (\ref{eq:IntegrabilityP}) now becomes
\bea
 \CQ'  + 6   \frac{ a'}{a} \CQ  +   \average{  \frac{\CR }{{\dot \Sc \strut}^2}   }'   +  2  \frac{a' }{a}  \average{  \frac{\CR}{{\dot \Sc \strut}^2} } + 4   \frac{ a'}{a} \CP    \nonumber   \\ 
\qquad \qquad = 16 \pi G \left( \average{  \frac{  \epsilon   }{{\dot \Sc \strut}^2}   }'  + 3  \frac{ a' }{a}     \average{  \frac{  \epsilon + p   }{{\dot \Sc \strut}^2} }      \right) + 2 \Lambda \average{\frac{1}{{\dot \Sc \strut}^{2}} }' ,
\eea
where the right-hand side obeys the averaged energy conservation law (\ref{eq:en_cons_averagedP}) that reduces to
\bea
\average{\frac{\epsilon}{{\dot \Sc \strut}^2} }' + 3 \frac{a'}{a} \average{\frac{\epsilon+p}{{\dot \Sc \strut}^2}} = - \average{\frac{\Theta}{{\dot \Sc \strut}^{\phantom{2\!\!}}} \, \frac{p}{{\dot \Sc \strut}^2}} + \average{\frac{\Theta + \dot \RatioU / \RatioU}{{\dot \Sc \strut}^{\phantom{2\!\!}}}} \average{\frac{p}{{\dot \Sc \strut}^2} } \nonumber \\
\label{eq:en_cons_averaged}
\qquad  + \average{\frac{\epsilon}{{\dot \Sc \strut}^2} \left( \frac{\dot \RatioU / \RatioU - 2 \ddot \Sc / \dot \Sc}{{\dot \Sc \strut}^{\phantom{2\!\!}}} \right) } - \average{ \frac{q^\mu a_\mu + \nabla_\mu q^\mu + \pi_{\mu \nu} \sigma^{\mu \nu}}{{\dot \Sc \strut}^3} } \, .\nonumber \\
\eea
{\bf Remark:} The requirement $\bm u \cdot \bm \nabla \Bound = 0$ in the choice of the window function corresponds to the definition of an averaging domain that follows the fluid flow. It thus ensures by construction the preservation over time of the collection of fluid elements to be averaged, in particular preserving their total rest mass (as shown in section \ref{massdef}) when it can be defined.

\subsubsection{Lagrangian window function:} 
The above equations for a comoving domain, $\bm u \cdot \bm \nabla \Bound = 0$,
simplify further when in addition the fluid proper volume measure is used, $\VecField = \bm u$, yielding a Lagrangian window function. This corresponds to setting $\RatioU = 1$ in equations (\ref{eq:HamiltonianAveraged})--(\ref{eq:en_cons_averaged}) above, dropping all terms that depend on its evolution. The system of averaged equations in the framework corresponding to the Lagrangian window function in \cite{foliations,generalfluid} is thus recovered, under an equivalent, here manifestly covariant form. As discussed in the above references, it becomes particularly transparent in a foliation by hypersurfaces of constant fluid proper time, $\Sc = \tau$.

\noindent
{\bf Remark:} The Lagrangian window function choice, based on a preferred fluid $4-$velocity field, is especially adapted to analyzing average properties within single-fluid cosmological models. This could apply, e.g. to the description of a dark matter-dominated late Universe within a dust model, or to the radiation-dominated era within a model of a pressure-supported fluid. It can as well be used in a model involving several non-comoving fluids, e.g. to describe a mixture of dark matter and radiation with different $4-$velocities. 
In this case, it would require choosing one of the fluids to be followed through its evolution and to define a proper volume measure. The total energy-momentum tensor would then have to be decomposed with respect to the corresponding frame, in which contributions from the other fluids will generally appear in the form of non-perfect fluid terms \cite{Ismael}.

\subsubsection{Riemannian volume averages:} As discussed at the end of section \ref{subsecAveragedEqsGeneral}, the choice of a Riemannian volume measure, $\VecField = \bm n$, is the most adapted for analyzing averaged geometric properties of the hypersurfaces themselves, e.g. by providing expressions of the averaged equations in terms of the extrinsic curvature of the hypersurfaces.  This is expected since the scale factor and averages are then based on the intrinsic spatial volume form of the hypersurfaces. 
The evolution equations for the scale factor with such a choice and for a comoving domain, $\bm u \cdot \bm \nabla \Bound = 0$, may be obtained from equations (\ref{eq:HamiltonianAveraged})--(\ref{eq:en_cons_averaged}) by setting $\Gamma = 1/\gamma$. This gives a manifestly covariant system of equations equivalent to that given in Appendix B of \cite{generalfluid}, also expressed in terms of the rest frame fluid variables. Recovering the dependence in the geometric variables such as the trace of extrinsic curvature then requires rewriting these local quantities along the lines suggested at the end of section \ref{subsecAveragedEqsGeneral}. 

\noindent
{\bf Remark:}
The choice of a Riemannian volume measure, $\VecField = \bm n$, is especially suited for studying the behaviour of hypersurfaces defined from geometric conditions, such as the Constant Mean Curvature requirement, which is frequently used in general relativity.

The averaged equations for this volume measure take their simplest form for a propagation of the domain along the normal vector $\bm n$ ($\bm n \cdot \bm \nabla \Bound = 0$). The evolution equations for such a choice of propagation of the domain can be directly obtained in terms of the geometric variables from the general equations of section \ref{subsecAveragedEqsGeneral}, recovering the framework and results of \cite{generalbackreac2}. 
However, a geometric propagation of the domain ($\bm n \cdot \bm \nabla \Bound = 0$) will in general imply a flow of fluid elements (with a $4-$velocity $\bm u$) across the domain boundary.
Preservation of fluid elements could be recovered with additional assumptions; for instance, for an irrotational fluid model with averaging defined in the corresponding global fluid rest frames, with $\bm n = \bm u$. In a more general cosmological setting, one may assume on large scales that vorticity effects may be neglected, at least near the domain boundary, allowing for a foliation where a propagation of the domain boundary along the normal vector would approximate a comoving propagation ($\bm u \cdot \bm \nabla \Bound = 0$). 
One may also assume a choice of hypersurfaces where statistical homogeneity holds for all observables, effectively leaving the evolution equations defined over such a choice of hypersurfaces invariant under the increase of scale of the domain $\Bound_0$ above a suitable homogeneity scale cut-off. This would then allow for a computation of averages over a global range ($\Bound_0 \longrightarrow + \infty$), effectively eliminating the need for distinguishing the possible propagations of the domain boundary for this choice; see \cite{rasanen} for an investigation of this framework.

\subsubsection{Light propagation:} As discussed in section \ref{WindowFunctionsLightcones}, an alternative choice for the domain boundary would be that of binding it to the past light cone of a given event by choosing the appropriate scalar $\Bound$ with light-like gradient, covering the evolution of the average properties of spatial sections in the interior of this light cone. 

Alternatively, one might consider the case where $\Sc$ has light-like gradient such that $\Sc = \Sc_0$ singles out a null surface that might be associated with the light cone of an observer, and where $\Bound$ has time-like or space-like gradient (e.g. $\bm \nabla \Bound$ being proportional to an irrotational fluid $4-$velocity $\bm u$).
Variation of average properties with respect to emitting times of the sources along a given cone then requires a variation of the parameter $\Bound_0$, while the above results for the dependence in $\Sc_0$ would provide insight on drift effects as the observer changes. 
These situations have been investigated in detail with similar covariant averaging schemes in \cite{lightconeav1}
(see also the application in an adapted coordinate system \cite{lightconeav2}).

\noindent
{\bf Remark:} Averaging domains defined from the light cone are natural candidates for relating the averaging formalism discussed in this paper to observations. 
It is important to keep in mind that the formalism presented in this paper is general, allowing for averaging over hypersurfaces of arbitrary globally hyperbolic space-times. In particular, the average equations only implicitly depend on the metric of space-time. While we consider this being an advantage, as it allows to express average properties independently of a specific form of the space-time metric, it implies the need for further specifications and assumptions in order to connect the general result to observations. 
For example, assumptions must be made in order to interpret averaged quantities defined over spatial hypersurfaces in terms of (averaged) energy, flux, etc., of photon bundles emitted by matter sources and absorbed by specified classes of observers. Such an interpretation may become more natural if the formalism is specified to light cone averaging \cite{lightconeav1}, but further assumptions would still be needed in order to close the system of averaged equations
(e.g. by specifying a model for the inhomogeneous metric \cite{lightconeav2}), and to relate the obtained averages to observational results that are usually based on idealizing assumptions on the geometry. It is beyond the scope of this paper to go into details about the difficult task of establishing connections between averaged cosmological evolution equations and (statistical) observations of selected observers.
For papers addressing the link between the averaging formalism and its observational interpretation, see e.g. \cite{larena,rasanen} (with a covariant formalism for global spatial averages in the second case), 
and \cite{nezihe,mikolaj} for local and bi-local investigations.

\section{Conclusion}
\label{Conclusion}

Covariance is a requirement for any physical theory, and a cornerstone in the formulation of general relativity. 
In this paper we have investigated scalar covariant formulations of global dynamics relevant for the description of backreaction effects in cosmology. 
We have considered a generalized window function, allowing for arbitrary foliation, spatial boundary, and volume measure.

We provided an explicitly covariant form for the commutation rule and for the spatially averaged scalar parts of Einstein's equations, with the associated integrability condition, using this general window function. The absence of restrictions imposed on the energy-momentum tensor of the fluid sources allows us to apply these schemes to the early Universe as well as to the matter-dominated later stages, and they cover all spatial scales down to which the fluid approximation can be considered as valid. Backreaction terms are introduced from these equations, and are thus also expressed under a manifestly covariant form. We then applied these results to the physically relevant subcase of a comoving domain. 

We have given a procedure for providing several possible decompositions of the commutation rule and the resulting averaged equations. This allows us, for example, to get rid of boundary terms, or to keep them as transparent boundary flux terms, for any choice of domain propagation. 
We have discussed the effect on averaged equations of a relabeling of the hypersurfaces in a given foliation, and we have stressed the importance of being able to physically interpret the chosen label. 

The formalism used in this paper provides a unifying framework encompassing various scalar averaging schemes that have been suggested or could be used for the description of averaged properties of cosmological models. It can be straightforwardly adapted to a given specific scheme by suitably choosing the window function. Several examples of such possible applications were given. In particular, we have shown that the manifestly covariant averaging scheme used in this work reduces to the averaging scheme considered in \cite{generalfluid} for a so-called Lagrangian window function, providing covariant formulas for the latter scheme. The explicit selection of the foliation by a scalar function in the scheme used in this work also makes it suitable for the forthcoming investigation of foliation dependence of averaged expressions \cite{foliationdependence}, and it may be helpful for other related considerations.

%------------------------------------------------
\ack
This work is part of a project that has received funding from the European Research Council (ERC) under the European Union's Horizon 2020 research and innovation programme (grant agreement ERC advanced grant 740021--ARTHUS, PI: TB). AH is supported by an University of Canterbury doctoral scholarship, and acknowledges hospitality for visits to CRAL--ENS, Lyon, supported by Catalyst grant CSG--UOC1603 administered by the Royal Society of New Zealand. 
Related work has been begun by AH at the Niels Bohr Institute under the supervision of Subir Sarkar. 
AH is also grateful for the support given by the funds: `Knud H{\o}jgaards Fond', `Torben og Alice Frimodts Fond', and `Max N{\o}rgaard og Hustru Magda N{\o}rgaards Fond'.
PM is supported by a `sp\'ecifique Normalien' Ph.D. grant, and acknowledges hospitality and support by Catalyst grant CSG--UOC1603 during a visit to the University of Canterbury. We thank L\'eo Brunswic, Mauro Carfora, Xavier Roy, Subir Sarkar, and David Wiltshire for valuable discussions, and the anonymous referees for useful comments. 
%------------------------------------------------

\appendix

\section{Freedom of hypersurfaces labeling}
\label{gauge1}

We here investigate in more detail the consequences of a change of the hypersurfaces label $\Sc$ (without change of the hypersurfaces) for the terms appearing in the evolution equations for the effective scale factor $a$.

Any transformation of the form 
\bea \label{eq:fMonotonic_trans}
\Sc \mapsto f(\Sc),
\eea
where $f$ is a strictly monotonically increasing function, is a transformation of the foliation of $\Sc$ onto itself (i.e. the same set of hypersurfaces is considered, with a different parametrization), since
\bea \label{eq:fMonotonic}
n_\mu = - \frac{\nabla_{\mu} f(\Sc)}{\sqrt{ -\nabla_{\nu} f(\Sc) \nabla^{\nu} f(\Sc) }} = - \frac{\nabla_{\mu} \Sc}{\sqrt{-\nabla_{\nu} \Sc \nabla^{\nu} \Sc }} \; . 
\eea
The class of transformations (\ref{eq:fMonotonic_trans}) is thus a gauge of the foliation. 

This seemingly innocent parametrization freedom can cause issues if we are naively evaluating averaged quantities without paying attention to the interpretation on what the time label $\Sc$ represents in the equations. 
As an example, the interpretation of the Friedmann equations under their usual form relies on the fact that their time parameter has a transparent meaning as the eigentime of ideal fundamental observers.

Let us consider an integrand 
\bea \label{eq:Integrand_gauge}
\Scalar \, W = - \Scalar \Vec^{\mu} \nabla_{\mu}( \heavy (\Sc_0 - \Sc) ) \,\heavy (\Bound_0 - \Bound) \; ,
\eea
where the vector $\Scalar \Vec^{\mu} \heavy (\Bound_0 - \Bound)$
is invariant under reparametrizations (\ref{eq:fMonotonic_trans}) of $\Sc$. (This is for instance the case if $\Scalar$, $\VecField$ and $\Bound, \Bound_0$ are independent of $\Sc$ or only depend on it \textit{via} the normal vector $\bm n$.)
Under such a reparametrization, the integral $I(\Scalar) = I(\Scalar)_{\Sc,\Sc_0}$ (recovering provisionally an explicit indication of the dependence in $\Sc$ and $\Sc_0$ of the window function) becomes
\bea \label{eq:I_trans_monotonic}
I (\Scalar)_{\Sc,\Sc_0}  \mapsto   I (\Scalar)_{f(\Sc),f(\Sc_0)} = I (\Scalar)_{\Sc,\Sc_0} \; , 
\eea
where we have used that
\bea \label{eq:heavy_monotonic}
\heavy(f(\Sc_0) - f(\Sc)) = \heavy(\Sc_0 - \Sc) \; ,
\eea
for strictly increasing functions $f$. 
Such an integral thus only depends on the chosen foliation and the selected slice, but not on the parametrization, and we can remove the subscript notation $\Sc,\Sc_0$ in the following.

Derivatives with respect to the parameter transform as
\bea \label{eq:I_trans_monotonic_deriv}
 \frac{   \partial I (\Scalar)}{  \partial \Sc_0 }  \mapsto    \frac{   \partial I (\Scalar) }{  \partial (f(\Sc_0))  } = \frac{1 }{f'(\Sc_0) }    \frac{ \partial  I (\Scalar)  }{  \partial \Sc_0  }  ,   
\eea
while second derivatives become
\bea \label{eq:I_trans_monotonic_deriv2}
 \frac{   \partial^2 I (\Scalar)}{  \partial \Sc_0 ^2 }  \mapsto  &  \frac{   \partial^2 I (\Scalar) }{  \partial (f(\Sc_0))^2 } = \frac{1}{f'(\Sc_0)^2} \,\frac{\partial^2 I(\Scalar)}{\partial \Sc_0^2} - \frac{f''(\Sc_0)}{f'(\Sc_0)^3} \,\frac{\partial I(\Scalar)}{\partial \Sc_0}  . 
\eea
We can therefore tune first derivatives by any positive rescaling $f'(\Sc_0)$ through the transformations (\ref{eq:fMonotonic_trans}), while second derivatives may even be canceled or change sign, since $f''(\Sc_0)$ is not constrained in its sign.
The above results similarly apply to the average $\average{\Scalar}$ and its derivatives with respect to $\Sc_0$. 

We conclude that, without a physical interpretation of the hypersurface label $\Sc$, statements about the magnitude of 
first-order derivatives (\ref{eq:I_trans_monotonic_deriv}), as well as \emph{any} statements (about magnitude or sign) about second-order derivatives (\ref{eq:I_trans_monotonic_deriv2}), are degenerate with the choice of $\Sc$. This applies for instance to the left-hand sides of the averaged dynamical equations (\ref{eq:HamiltonianAveragedP})--(\ref{eq:RaychaudhuriAveragedP}), or (\ref{eq:HamiltonianAveraged})--(\ref{eq:RaychaudhuriAveraged}), that are proportional to $(\partial I(1)/\partial \Sc_0)^2$ and $\partial^2 I(1) / \partial \Sc_0^2$, assuming that $\VecField$, $\Bound$ and $\Bound_0$ are defined independently of $\Sc$ or only depend on it \textit{via} the normal vector $\bm n$.

Under the same assumption, the conclusions about parametrization-dependence also hold for the terms on the right-hand sides of (\ref{eq:HamiltonianAveragedP})--(\ref{eq:RaychaudhuriAveragedP}). Most of them can be written as
$\average{ \,\Scalar / (P^\sigma \nabla_\sigma \Sc)^2 \,}$
with a scalar $\Scalar$ that is unchanged under the reparametrization (\ref{eq:fMonotonic_trans}), even when it depends on $\Sc$, such as $\Ratio$, and would thus rescale by a factor $f'(\Sc_0)^2$, as does $(\partial I(1) / \partial \Sc_0)^2$. The only exception is the combination of terms $\average{ - (\Theta_P + \Ratio^{-1} P^\mu \nabla_\mu \Ratio) \, P^\nu \nabla_\nu (P^\sigma \nabla_\sigma \Sc) \, (P^\rho \nabla_\rho \Sc)^{-3} }$ appearing in $\CP$ in (\ref{eq:RaychaudhuriAveragedP}), which would transform as
\bea
\, \average{ - \frac{(\Theta_P + \Ratio^{-1} P^\mu \nabla_\mu \Ratio)  \, P^\nu \nabla_\nu (P^\sigma \nabla_\sigma \Sc)}{(P^\rho \nabla_\rho \Sc)^{3}} } \mapsto \nonumber \\
\frac{1}{f'(\Sc_0)^2} \average{ - \frac{(\Theta_P + \Ratio^{-1} P^\mu \nabla_\mu \Ratio)  \, P^\nu \nabla_\nu (P^\sigma \nabla_\sigma \Sc)}{(P^\rho \nabla_\rho \Sc)^{3}} }   - \frac{f''(\Sc_0)}{f'(\Sc_0)^3} \frac{\partial I(1)}{\partial \Sc_0} \; ,
\eea
i.e. in the same way as $\partial^2 I(1) / \partial \Sc_0^2$.
These identical transformations of both sides of the averaged evolution equations ensure the preservation of the form of these equations under a reparametrization. The same remarks hold for the equations (\ref{eq:HamiltonianAveraged})--(\ref{eq:RaychaudhuriAveraged}) with $\bm P = \bm u$ in this case.

% ==================================================
% ==================================================
% BIBLIO

\end{document}